\documentclass{article}
\usepackage{setspace}
\usepackage{arxiv}

\usepackage[utf8]{inputenc} 
\usepackage[T1]{fontenc}    
\usepackage{url}            
\usepackage{booktabs}       
\usepackage{amsfonts}       
\usepackage{nicefrac}       
\usepackage{microtype}      
\usepackage{float}          
\usepackage{graphicx}
\usepackage[round,authoryear]{natbib}
\usepackage{hyperref}
\usepackage{doi}
\usepackage{subfiles}
\usepackage{amsmath}
\usepackage{bm} 
\usepackage{subcaption}
\usepackage{xcolor}
\usepackage{ulem}

\usepackage[mathlines]{lineno}

\title{Variability in Performance of a Machine-Learning Seismicity Catalog: Central Italy, 2016–2017}

\author{%
  \href{https://orcid.org/0000-0003-2960-4601}{\includegraphics[scale=0.06]{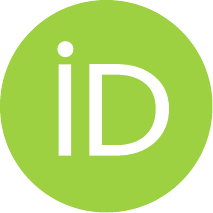}\hspace{1mm}Jaehong Chung}\textsuperscript{1} \quad
  \href{https://orcid.org/0000-0001-5986-9702}
  {\includegraphics[scale=0.06]{orcid.pdf}\hspace{1mm}Yifan Yu}\textsuperscript{1} \quad
  \href{https://orcid.org/0000-0002-9697-6504}
  {\includegraphics[scale=0.06]{orcid.pdf}\hspace{1mm}Lauro Chiaraluce}\textsuperscript{2} \quad
  \href{https://orcid.org/0000-0001-9039-3503}
  {\includegraphics[scale=0.06]{orcid.pdf}\hspace{1mm}Maddalena Michele}\textsuperscript{2}
  \href{https://orcid.org/0000-0002-8667-1838}
  {\includegraphics[scale=0.06]{orcid.pdf}\hspace{1mm}Gregory C. Beroza}\textsuperscript{1} \quad  
  \\[8pt] 
  \textsuperscript{1}Department of Geophysics, Stanford University, Stanford, CA, USA  \\
  \textsuperscript{2}Istituto Nazionale di Geofisica e Vulcanologia, Italy  \\  
  [3pt]  %
}

\date{}

\renewcommand{\undertitle}{ }
\renewcommand{\shorttitle}{ } 

\hypersetup{
pdftitle={A template for the arxiv style},
pdfsubject={q-bio.NC, q-bio.QM},
pdfauthor={David S.~Hippocampus, Elias D.~Striatum},
pdfkeywords={First keyword, Second keyword, More},
}

\begin{document}
\maketitle

\begin{abstract}
Machine learning (ML) catalogs contain many more earthquakes than routine catalogs, but their performance in phase picking and earthquake detection has not been fully evaluated. We develop station-level detection probabilities using logistic regression and combine them across a seismic network to compute spatial magnitude-of-completeness fields. We apply this approach to two catalogs from the 2016–2017 Central Italy sequence that were constructed from the same seismic network, one routine and one ML based. At the station level, the ML picker increases detection sensitivity by identifying smaller magnitude events and detecting earthquakes at greater distances. Spatially, the magnitude-of-completeness decreases substantially, with median values shifting from 1.6 to 0.5 for P waves and from 1.7 to 0.5 for S waves. However, the ML catalog also shows greater variability in station-level performance than the routine catalog. These results demonstrate that ML-based improvements in detectability are widespread but spatially non-uniform, highlighting their benefits, their limitations, and the potential for further improvements. 
\end{abstract}


\section{Introduction}\label{sec:intro}
Deep-learning methods have become a useful tool for generating high-resolution earthquake catalogs. Machine learning (ML) phase pickers trained on large volumes of labeled seismograms and association algorithms can identify many small earthquakes and provide consistent P- and S-wave arrival times across large datasets \citep{perol2018conv, ross2018generalized, zhu2019phasenet, mousavi2020earthquake, mcbrearty2023earthquake}. The resulting ML-based catalogs generally exhibit lower overall completeness magnitudes, $M_c$, and have facilitated new and clearer observations of fault-zone structure, fluid migration, and volcanic unrest \citep{tan2021machine, beroza2021machine, gong2023machine, kato2024implications, suzuki2025forearc, tan2025clearer}.

Despite these advantages, the performance of ML-based monitoring is not uniform. Detection accuracy varies with noise conditions, station geometry, and signal characteristics that differ from the training distribution \citep{zhong2024deep}, and continuous processing reveals sensitivity to workflow choices and prediction consistency \citep{park2023mitigation, pita2023parametric}. Although many studies report a lower catalog-wide $M_c$, such scalar estimates may obscure strong spatial variability in detectability \citep{noel2025challenges} In practice, completeness depends on station coverage, station characteristics, hypocentral distance, attenuation, and frequency content, and can remain limited in regions with sparse instrumentation. The need for spatially resolved detectability has long been recognized for reliable monitoring, early warning, and network design \citep{kijko1977algorithm_I, kijko1977algorithm_II, schorlemmer2008probability, mignan2011bayesian}. The detection capability of modern ML-based catalogs has not been systematically quantified so that the extent to which ML enhances detectability remains unclear.

A spatial perspective helps identify where a catalog is most complete, where ML-based methods provide meaningful gains, and where further improvement may require retraining or additional stations. Such information is essential for evaluating catalog quality, guiding network expansion, and linking monitoring performance to seismic-risk assessments \citep{keil2023optimal}. No previous study has provided a quantitative, data-driven comparison of spatial detectability between ML-based and routine catalogs.

We address this gap using a probability-based magnitude of completeness (PMC) framework that converts station-level detection statistics into spatial magnitude-of-completeness maps. Building on prior work on network detectability and spatial completeness \citep{schorlemmer2008probability, mignan2011bayesian}, we estimate station-level detection probabilities as a function of magnitude and distance, combine them across the seismic network, and determine spatial $M_c (\bm{x})$ using a high-confidence criterion. We apply this framework to two parallel catalogs from the 2016–2017 Central Italy sequence—one catalog generated using ML-based phase picking, and the other catalog produced using routine procedures—and quantify where and by how much ML improves spatial detectability for P- and S-wave arrivals. The results show broad reductions in $M_c (\bm{x})$ while also revealing regions where improvements are limited. This spatial view clarifies the local detectability changes introduced by the ML-based catalog and highlights where these changes are most substantial or only modest across the network. We also find that the variability of performance of ML-based phase picking for different stations is greater than that for conventional phase picking. This suggests that substantial further improvements to machine learning catalogs should be possible. 

\section{Data: routine vs. machine learning catalogs for the 2016-2017 Central Italy sequence}\label{sec:data}
The 2016–2017 Central Italy seismic sequence consisted of multiple large earthquakes, including the $M_w$ 6.0 Amatrice event on 24 August 2016 and the $M_w$ 6.5 Norcia event on 30 October 2016, which ruptured an extended portion of the central Apennines \citep{chiaraluce2022comprehensive, scognamiglio2018complex, chiarabba2018faults}. Because of the high societal impact, strong spatial heterogeneity in aftershock activity, and the need for rapid situational awareness, this sequence was monitored with exceptional detail. The Istituto Nazionale di Geofisica e Vulcanologia (INGV) augmented its permanent network with a dense set of temporary stations deployed during the emergency response \citep{moretti2017sismiko}. The combination of intense aftershock activity and dense instrumentation enabled the development of both conventional and machine-learning–based catalogs, providing an important opportunity to evaluate differences in detection capability. 

\begin{figure}
    \centering
    \includegraphics[width=1.0\textwidth]{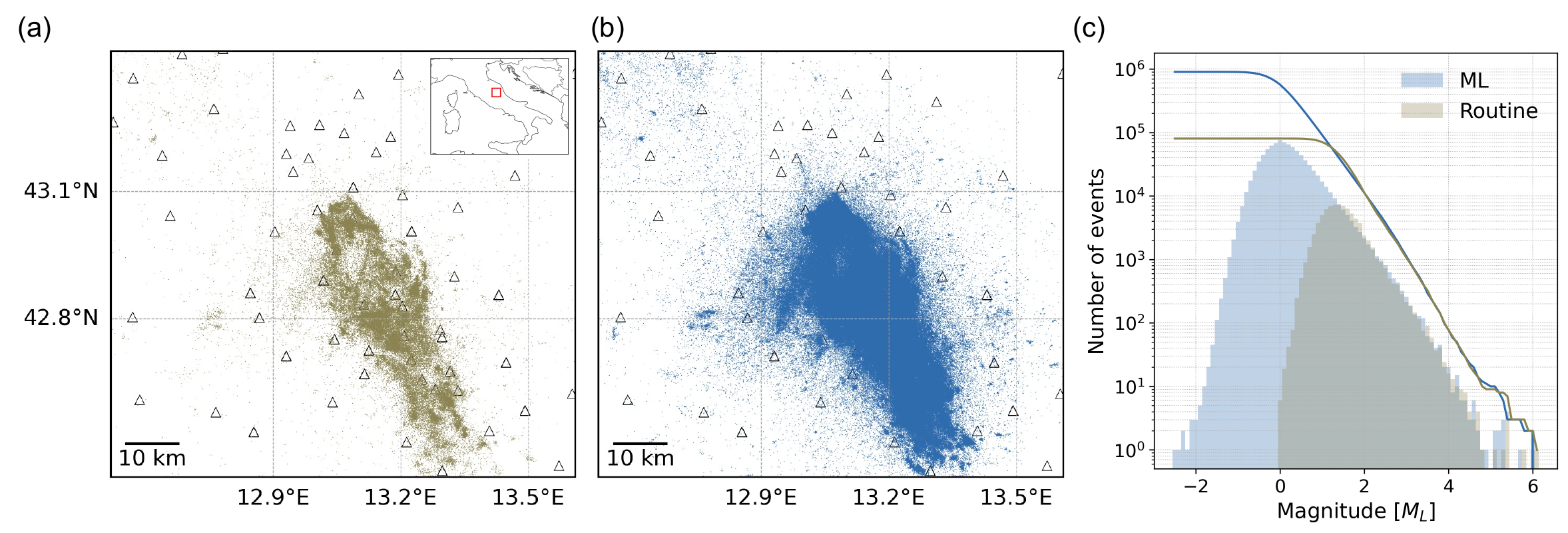}
    \caption{Earthquake catalog data: 
    (a) Map of the study area showing earthquakes from the routine catalog (olive) together with stations of the INGV network (triangles).
    (b) Same region, plotted for events in the ML catalog (blue).
    (c) Magnitude–frequency distributions for both catalogs, shown as histograms (filled) and cumulative curves for local magnitude ($M_L$).
    A bin width of 0.1 magnitude units is used for both datasets.
    Previously reported global completeness magnitudes are $M_c = 1.6$ for the routine catalog and $M_c = 0.2$ for the ML-based catalog \citep{chiaraluce2022comprehensive}.}
    \label{fig:catalog_summary}
\end{figure}

In this study, we use two parallel catalogs covering August 2016 to August 2017 as shown in Figure \ref{fig:catalog_summary}. The routine catalog (CAT1 in \citep{chiaraluce2022comprehensive}) contains 82,356 earthquakes with $0.0 \leq M_L \leq 6.12$, obtained by relocating the real time routine catalog (CAT0) arrivals with an expanded station set that includes temporary stations installed after the sequence onset. The machine-learning catalog includes 900,050 earthquakes with $-2.6 \leq M_L \leq 6.1$, obtained by applying the PhaseNet neural-network picker \citep{zhu2019phasenet} and automatic association and relocation procedures \citep{tan2021machine}. In particular, for a fair comparison, we evaluate both catalogs using the same set of stations: the INGV permanent network together with the temporary stations deployed immediately after the Amatrice mainshock \citep{moretti2017sismiko}. We exclude phase picks from additional stations deployed later by the British Geological Survey and the University of Edinburgh that were included in the ML workflow \citep{chiarabba2018faults} to avoid bias from differences in station coverage. Further details about the catalogs are provided in \citet{chiaraluce2022comprehensive}.

\section{Methods}\label{sec:method}
\begin{figure}
    \centering
    \includegraphics[width=1.0\linewidth]{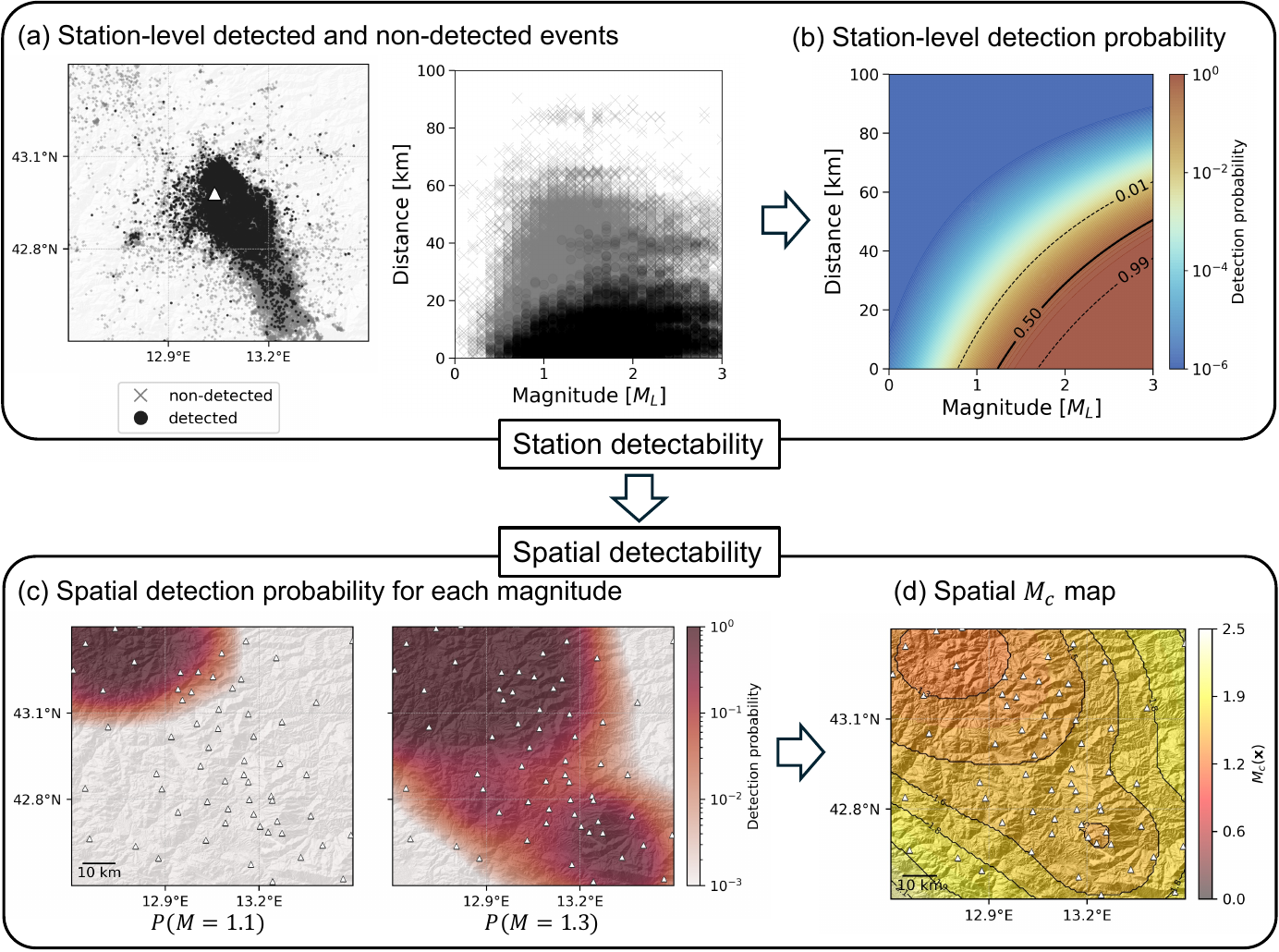}
    \caption{
    Workflow for quantifying spatial, probability-based magnitude of completeness: (a) Station-level detected and non-detected P-wave picks for an example station, FEMA, in the IV network, shown in both map view and magnitude–distance space, (b) Station-level detection probability calculated from the observations in (a). (c) Spatial detection probability for selected magnitudes; examples shown for $M=1.1$ and $M=1.3$. (d) Resulting spatial magnitude of completeness map ($M_c (\bm{x})$).}
    \label{fig:workflow}
\end{figure}

We follow the probability-based magnitude of completeness (PMC) framework proposed by \citet{schorlemmer2008probability}, which combines station-level detection capability into a network-level detection probability and estimates the magnitude at which the network reliably detects events. We extend and adjust this framework in three ways: we introduce a continuous station-level detection model that removes abrupt changes associated with magnitude–distance binning; we quantify detection probabilities separately for P- and S-wave arrivals; and we apply a stricter network criterion that requires at least eight detecting stations rather than the minimum four. Figure \ref{fig:workflow} shows the overall workflow incorporated in the station-level and network-level formulation described below.

\subsection{Station-level detection probability}
We compile station-level detection information from the catalog by assigning a binary label indicating whether a station recorded a P- or S-wave arrival for each event. We examine detection behavior as a function of event magnitude and hypocentral distance, which we compute from each event’s latitude, longitude, and a median ML-based catalog depth of 5.475~km following \citet{schorlemmer2010completeness}. This produces a two-dimensional distribution of detections and non-detections in magnitude–distance space for each station.

In the approach of \citet{schorlemmer2008probability}, station-level detection probabilities were estimated by counting events within magnitude–distance bins. Because the estimates depend on bin size and the number of events within each bin, this approach can produce abrupt probability changes in sparsely sampled regions, even when assuming that detection probability increases monotonically with magnitude (see Figure S1). To obtain a smooth and continuous probability surface directly from the data, we instead develop a logistic-regression–based station-level detection model.

For each station $i$ and phase type $\phi \in \{P, S\}$, we define the shifted magnitude and the station-level detection probability $p_{i}^{\phi}$ as follows:

\begin{align}
    \mathbf{M}^* &= \mathbf{M} - M_\text{min}, \\
    p_{i}^{\phi}(\mathbf{M}^*, \mathbf{L}_i)
    &= \frac{1}{1 + \exp\!\left[-\left(
    \alpha_i^{\phi} + \beta_i^{\phi}\,\mathbf{M}^*
    + \gamma_{i}^{\phi}\,\mathbf{L}_i
    + \eta_{i}^{\phi} (\mathbf{M}^* \odot \mathbf{L}_i) 
    \right)\right]} ,
\end{align}

where $\mathbf{M}$ is the vector of event local magnitudes, $M_\text{min}$ is the smallest magnitude in the corresponding catalog (routine or ML), $\mathbf{L}_i$ is the vector of hypocentral distances between events and station~$i$, the operator $\odot$ denotes the Hadamard product (element-wise multiplication), and $\alpha_i^\theta$, $\beta_i^\theta$, $\gamma_i^\phi$, and $\eta_i^\phi$ are station- and phase-specific fitting paramters. To ensure a consistent probability model within the data-supported domain ($M>M_\text{min}$ and $0\leq L \leq 150$ km), we constrain the model so that detection probability decays toward zero by the maximum distance and is set to zero for magnitudes below the catalog minimum.

\subsection{Spatial detection probability and magnitude of completeness}
We compute the network-level spatial detection probability by combining the station-level detection probabilities $p_{i}^\phi(\mathbf{M}^*,\mathbf{L}_i)$ under the conditional-independence framework of \citet{schorlemmer2008probability}. At each location, we require detections at a minimum of eight stations ($N \ge 8$), reflecting a conservative figure for the number of independent phase picks needed to obtain stable hypocentral constraints in relocation and double-difference workflows \citep{gomberg1990effect, waldhauser2001hypodd}. To quantify this, we use the complement rule:

\begin{equation}
P(\mathbf{x}\mid M, N\ge 8)
=
1 - \sum_{n=0}^{7} P(\mathbf{x}\mid M, N=n),
\end{equation}

where each term $P(\mathbf{x}\mid M, N=n)$ is computed by summing over all unordered $n$-station subsets, multiplying the detection probabilities for the selected stations and the non-detection probabilities for the remaining stations.

We determine the spatial magnitude of completeness $M_c(\mathbf{x})$ as the smallest magnitude for which the network achieves a high detection probability. Following \citet{schorlemmer2008probability}, we use a conservative threshold $P_c=0.99999$ and define

\begin{equation}
M_c(\mathbf{x})
=
\arg\min_M
\left\{
M \,\middle|\,
P(\mathbf{x}\mid M,\,N\ge 8) \ge P_c
\right\}.
\end{equation}

This yields a spatial field representing the smallest magnitude that the network can reliably detect at each location.

\section{Results and Discussion}\label{sec:resultsANDdiscussion}
\subsection{Station-level Detectability Differences}
We estimate station-level detection probabilities using a logistic-regression model that relates detection to event magnitude and hypocentral distance. These station-level probabilities are then combined across the network using a combinatorial joint-probability formulation to obtain spatial detection probabilities and the resulting spatial magnitude of completeness (Figure~\ref{fig:workflow}; details in Section~\ref{sec:method}).

\begin{figure}
    \centering
    \includegraphics[width=0.95\linewidth]{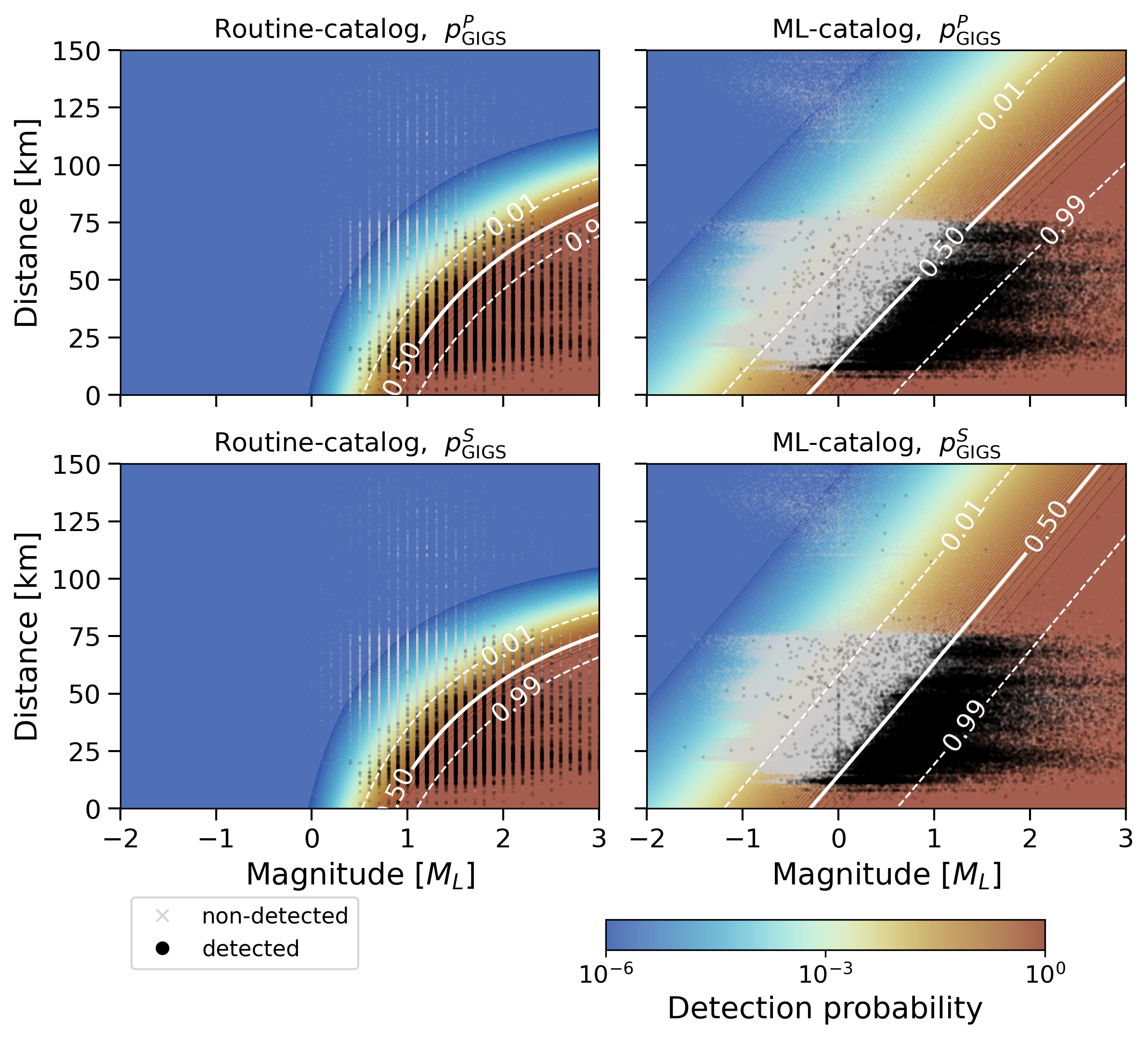}   
    \caption{Derived detection probability for station GIGS in the IV network, with detected (black dots) and non-detected (gray crosses) events. The left column shows the routine catalog and the right column the ML-based catalog; the top row is for P waves and the bottom row for S waves. Additional examples from other stations are provided in Figures S2—S5.
    }
    \label{fig:GIGS-station-level}
\end{figure}

We quantify station-level detectability for each phase and catalog.
Figure \ref{fig:GIGS-station-level} shows the derived detection probabilities based on the magnitude–distance distribution of arrivals at station GIGS. The routine catalog contains 17,027 detected P-wave arrivals and 12,163 detected S-wave arrivals, whereas the ML picker identifies 44,447 P-wave and 47,888 S-wave arrivals, corresponding to increases by factors of 2.6 and 3.9 for P and S waves, respectively.

These differences directly affect the estimated detection probabilities. At 50~km, the magnitude at which detection probability reaches 50~\% decreases from 1.69 (routine) to 0.82 (ML) for P waves, and from 1.79 (routine) to 0.73 (ML) for S waves. At a fixed magnitude of $M_L = 1.0$, the distance at which detection probability reaches 50~\% increases from 14.0~km (routine) to 57.6~km (ML) for P waves, and from 14.0~km (routine) to 64.3~km (ML) for S waves. The magnitude interval between 1~\% and 99~\% detection probability also widens in the ML models, from 0.87 to 1.84 magnitude units for P waves and from 0.92 to 1.79 for S waves, indicating a more gradual decrease in detection probability with magnitude and distance while still enabling detection of a population of smaller and more distant events. The distinct shapes of these detection probability surfaces, curved for routine picking versus more linear for ML picking, imply that their detection thresholds may be governed by different physical mechanisms. This poses the question of what leads to this difference in behavior. The linear boundary observed in the ML catalog suggests that detection is primarily limited by amplitude attenuation of coherent energy. Based on the ground motion attenuation relation $A\propto10^MR^{-1}\exp{-\pi fR/Qv}$, where $Q$ is the quality factor and $v$ is the seismic velocity, the detection threshold in terms of magnitude $M$ and distance $R$ scales as $M\propto \log(R) + \frac{\pi f}{Qv\ln{10}}R$. Considering only the attenuation term, the detection threshold follows the linear scaling: $M\propto \frac{\pi f}{Qv\ln{10}}R$. The routine catalog exhibits a distance saturation pattern that flattens with increasing magnitude. This may be explained by the decay of the high frequency content of the waveform, for which attenuation removes the high-frequency energy and suppresses the impulsive onset required by conventional pickers. This distinction can also be quantified using the slope of the detection probability boundary, $dR/dM \approx \left[\frac{\ln{10} \,Qv}{\pi}\right]f^{-1}$. For the routine catalog, the observed slope is approximately 20 km per unit magnitude, implying a dominant picking frequency of 10–15 Hz, which is rapidly attenuated. In contrast, the ML catalog exhibits a significantly steeper slope, indicating that the deep learning models effectively utilize amplitude information across a broader range of frequencies that remain coherent at greater distances.

\begin{figure}
    \centering
    \includegraphics[width=0.8\linewidth]{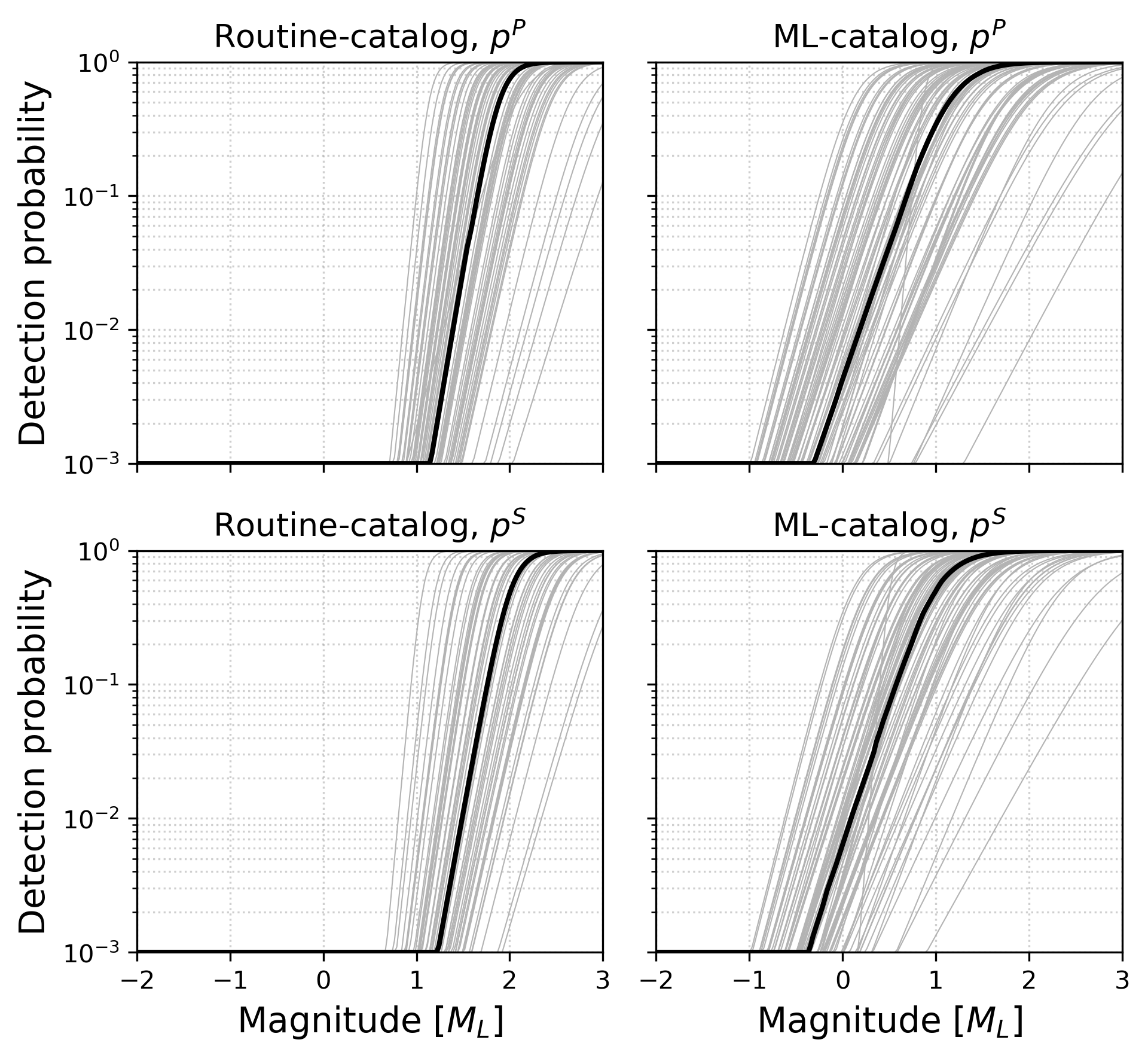}   
    \caption{Station-level detection probability as a function of magnitude at a fixed distance of 50~km. The left column shows the routine catalog and the right column the ML-based catalog; the top row shows P waves and the bottom row S waves. Thick curves show the median detection probability for each catalog and phase, and thin lines show individual-station curves.}    
    \label{fig:Median-station-level-magnitude}
\end{figure}

\begin{figure}
    \centering
    \includegraphics[width=0.8\linewidth]{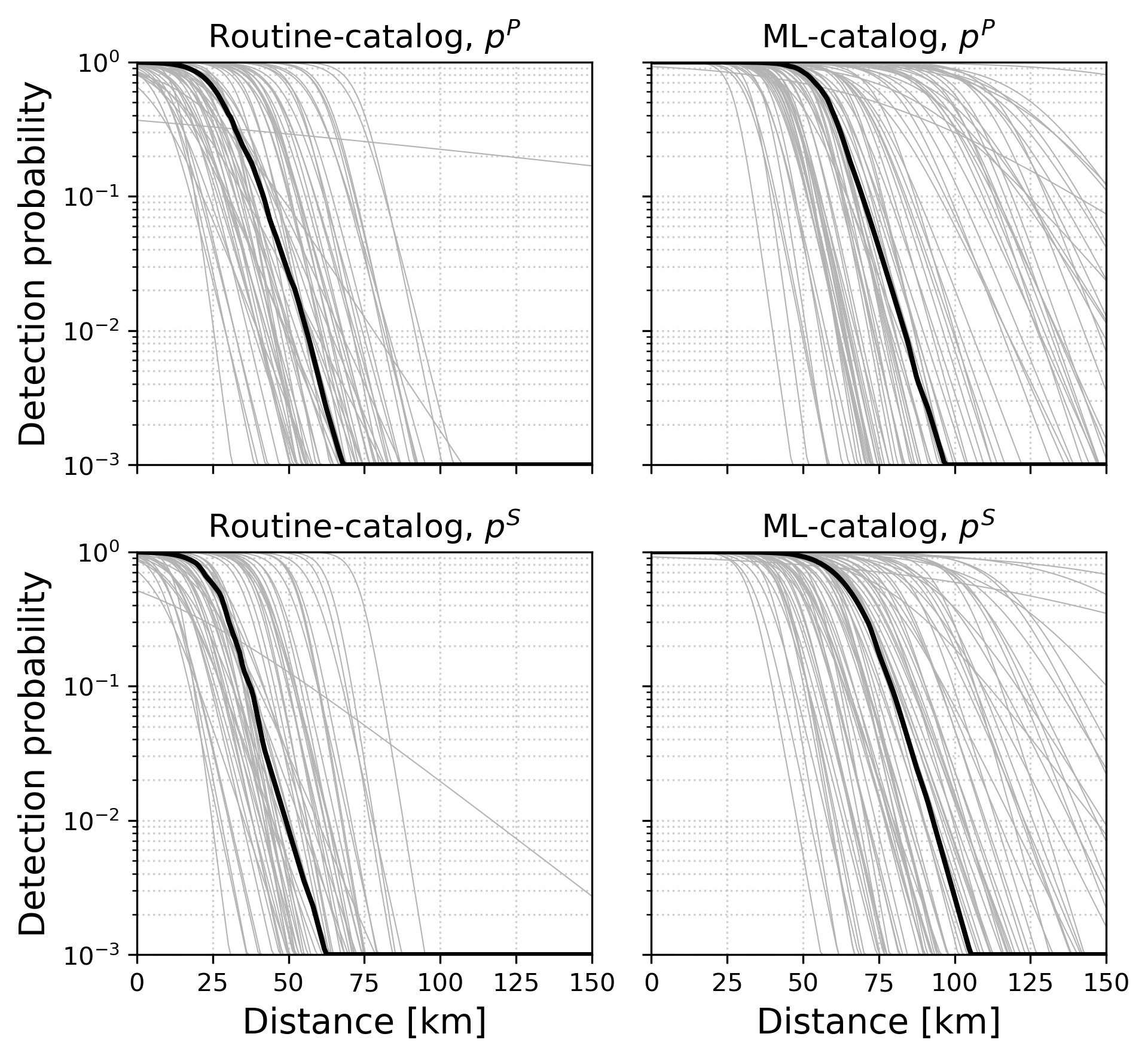}   
    \caption{Station-level detection probability as a function of distance at a fixed magnitude of $M_L = 1.5$. The left column shows the routine catalog and the right column the ML-based catalog; the top row shows P waves and the bottom row S waves. Thick curves show the median detection probability for each catalog and phase, and thin lines show individual-station curves.}
    \label{fig:Median-station-level-distance}
\end{figure}

Figure~\ref{fig:Median-station-level-magnitude} and \ref{fig:Median-station-level-distance} summarize these station-level changes across the full network. Using the network-median detection curves in Figure \ref{fig:Median-station-level-magnitude}, the magnitude at which detection probability reaches 50~\% at a distance of 50~km decreases from $M_{50}=2.26$ in the routine P-wave model to $M_{50}=1.73$ in the ML model. For S waves, the corresponding values decrease from $M_{50}=3.10$ (routine) to $M_{50}=1.47$ (ML). At a fixed magnitude of $M_L = 1.5$, the distance at which the median network curve reaches 50~\% detection probability increases from 16.4~km (routine P) to 46.5~km (ML P), and from 10.5~km (routine S) to 50.7~km (ML S) (Figure \ref{fig:Median-station-level-distance}). These station-level comparisons show that, for the same network and stations, the ML picker reduces the magnitude required for 50~\% detection by about 0.5~units for P waves and about 1.6~units for S waves, and allows $M_L \approx 1.5$ events to be detected roughly 30~km farther for P waves and more than 40~km farther for S waves compared to the routine picker. This pronounced improvement in S-wave detectability likely reflects the ML picker's ability to identify S-wave arrivals obscured by the P-wave coda, which is more challenging for amplitude-based detection methods \citep{zhu2019phasenet}. 

Although the ML-based approach improves station-level detectability compared to the routine picker in most respects, it also exhibits greater variability across stations. In particular, the standard deviation of station-level $M_{50}$ increases from 0.45 (routine P) to 0.65 (ML P), and from 0.46 (routine S) to 0.59 (ML S). Similarly, the standard deviation of station-level $R_{50}$ increases from 16.7~km (routine P) to 24.8~km (ML P), and from 15.8~km (routine S) to 23.1~km (ML S). The station-to-station variability in the performance of the machine learning model may reflect differences in noise conditions, waveform characteristics, or site effects, which can motivate retraining or fine-tuning of the neural network phase picker \citep{tan2025clearer}. It suggests that there exists the potential for substantial future improvements in the performance of machine learning models. These topics will be further investigated in future work.

\subsection{Spatial detection evolution}

\begin{figure}
    \centering
    \includegraphics[width=1.0\linewidth]{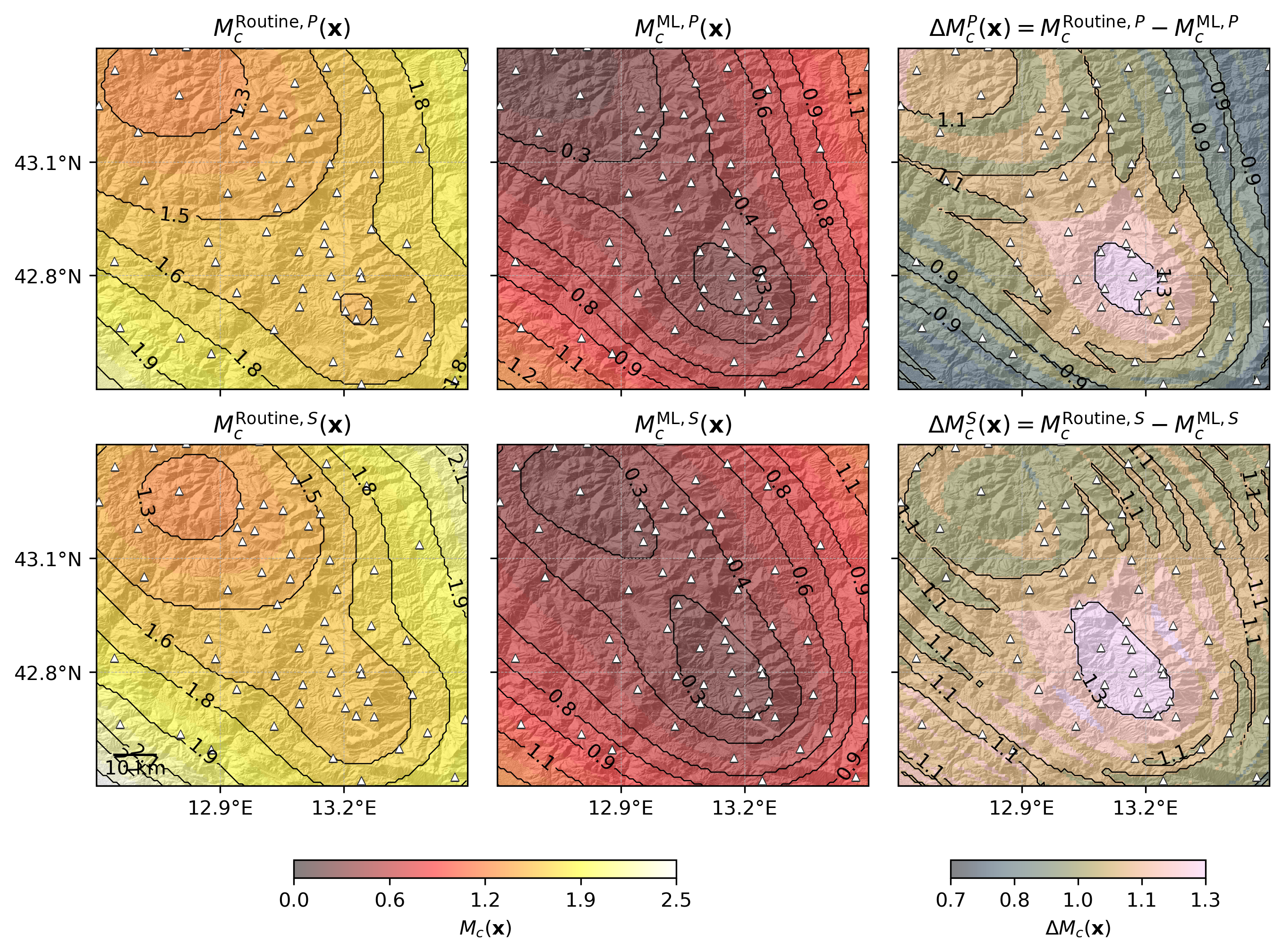}
    \caption{Spatially varying magnitude of completeness $M_c(\mathbf{x})$ for P- and S-wave detection. 
    The left column shows $M_c(\mathbf{x})$ estimated from the routine catalog, the middle column shows 
    $M_c(\mathbf{x})$ from the ML-based catalog, and the right column shows their difference 
    $\Delta M_c(\mathbf{x}) = M_c^{\mathrm{Routine}} - M_c^{\mathrm{ML}}$. 
    Positive values in $\Delta M_c(\mathbf{x})$ indicate regions where the ML picker lowers the detection threshold relative to the routine catalog. 
    Triangles denote station locations. 
    }
    \label{fig:mc_spatial}
\end{figure}

We assess network-level changes in detection capability using the spatial magnitude of completeness, $M_c(\bm{x})$, derived for both catalogs and phases (Figure~\ref{fig:mc_spatial}). The routine P-wave completeness map has a median value of 1.6; whereas, the ML-based map shifts to a median of 0.5, indicating a consistent increase of detection sensitivity across the region. The median reduction in magnitude of completeness is 1.0 magnitude unit for P-waves and 1.1 for S-waves, and every grid point shows a positive reduction. In particular, reductions greater than one magnitude unit occur over 41~\% of the study area for P-waves and 71~\% for S-waves. The largest decreases occur in areas with denser station coverage along the aftershock zone (approximately 13.2 $^\circ$E, 42.8$^\circ$N); whereas, regions near the network perimeter show relatively moderate but still positive changes. The trend is that $M_c$ inversely correlates with local station density \citep{Nanjo2010}. The $M_c$ from the ML catalog in the region with dense station coverage is much lower than previous studies based on the routine catalog \citep{wiemer2000minimum, gaebler_performance_2021}. We note that the detection capability of the seismic network in our analysis is not a comprehensive measurement of the performance of the network, as reliable phase picking is necessary, but not sufficient for accurate earthquake locations, which are one of the main goals of seismic monitoring \citep{Alessandro2011}. 

\begin{figure}
    \centering
    \hfill
    \begin{subfigure}[b]{0.4\textwidth}
        \centering
        \includegraphics[width=\linewidth]{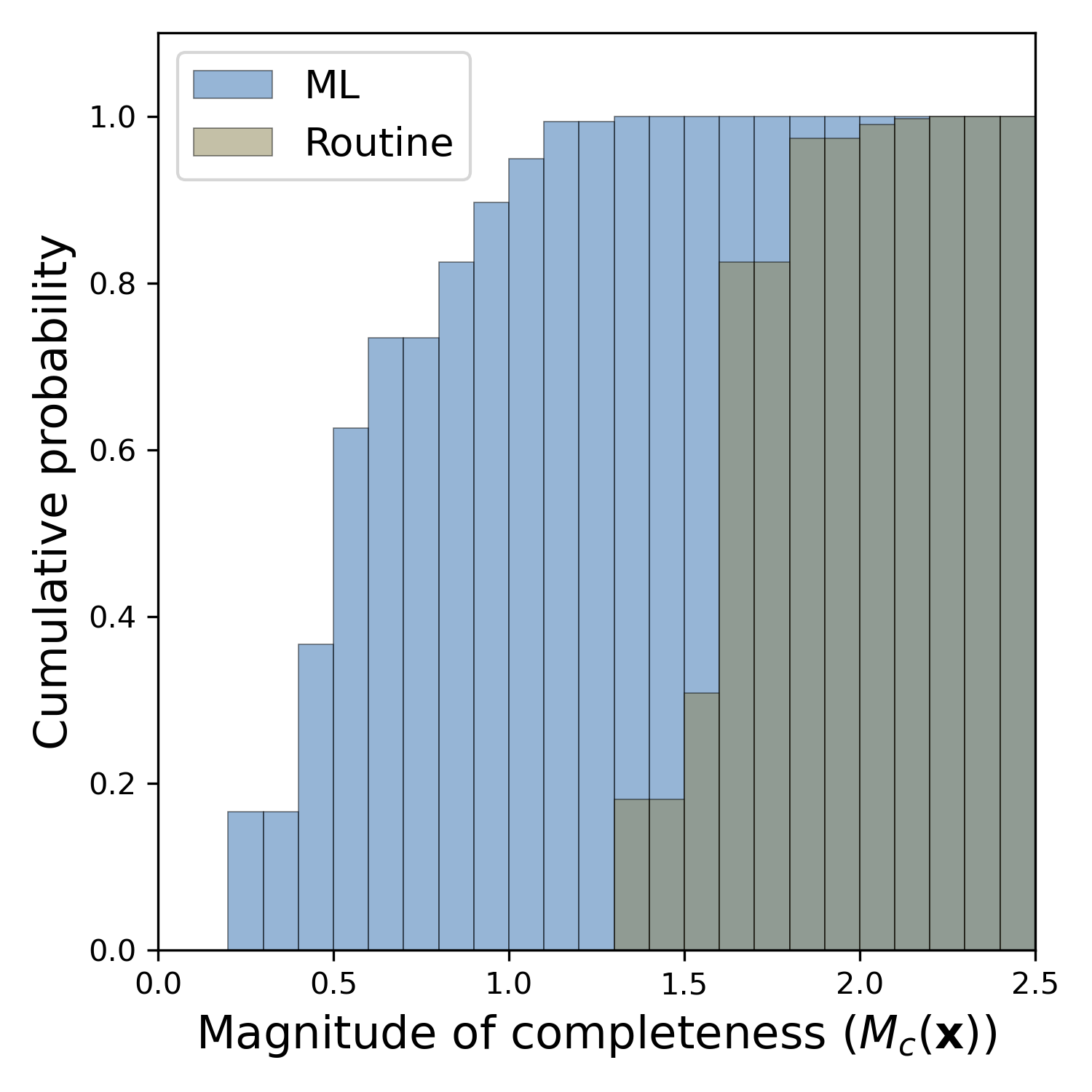}
    \end{subfigure}
    \hfill
    \begin{subfigure}[b]{0.4\textwidth}
        \centering
        \includegraphics[width=\linewidth]{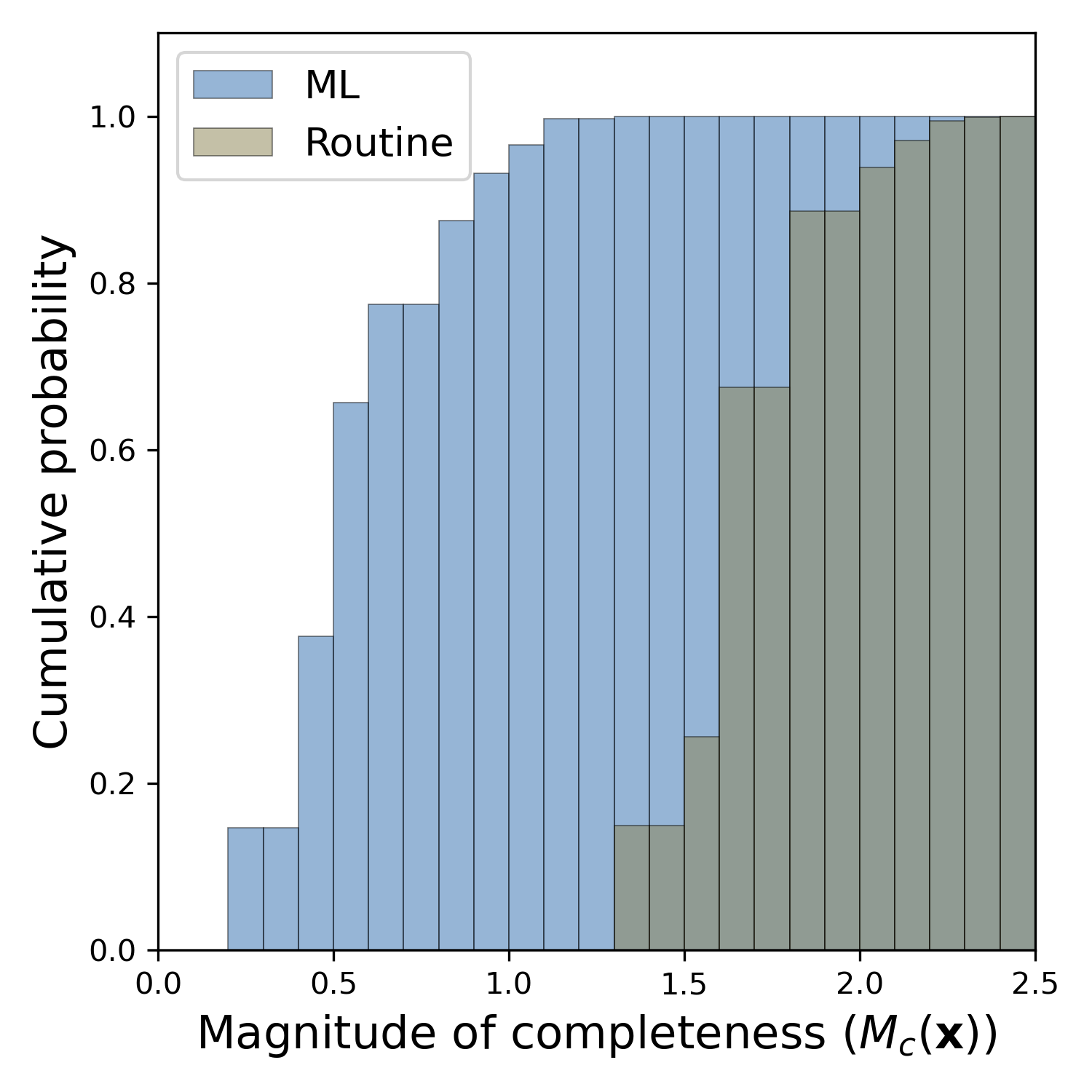}
    \end{subfigure}
    \hfill

    \caption{Cumulative distributions of $M_c(\mathbf{x})$ for each catalog,
    shown separately for (left) P waves and (right) S waves.}
    \label{fig:mc_spatial-histogram}
\end{figure}

The cumulative distributions in Figure \ref{fig:mc_spatial-histogram} quantify these spatial variations. For P-waves, the ML-based completeness values span 0.2–1.4 with a median of 0.5, whereas the routine catalog spans 1.3–2.2 with a median of 1.6. For S-waves, the ML catalog ranges from 0.3–1.4 with a median of 0.5, compared to 1.3–2.6 and a median of 1.7 for the routine catalog. In both phases, the upper end of the ML distributions (approximately 1.3–1.4) coincides with the lower end of the routine distributions, and the two sets of curves do not overlap elsewhere. These contrasts demonstrate that the ML picker produces consistently lower completeness values across the region, while still capturing spatial variability in $M_c(\mathbf{x})$.

\subsection{Potential and Limitations of the ML Catalog and the PMC Framework}
We used a probability-based magnitude of completeness (PMC) framework to quantify detection performance at the station level and to propagate these estimates into spatially continuous magnitude of completeness fields. This provides a consistent approach to compare detection levels between catalogs, to identify where monitoring performance is limited, and to evaluate how station-level sensitivity translates into network-level detectability. The framework is flexible and can be used to assess seismic monitoring performance and to highlight areas where additional station coverage, or improved picking methods, would most effectively improve detection thresholds. It can also be extended to estimate the detection probability density function over separate pre-, co-, and post-seismic intervals, enabling time-dependent assessment of network performance as noise levels vary throughout the sequence.

When applied to the 2016–2017 Central Italy sequence, the framework shows that the machine learning based catalog produces systematically lower magnitude of completeness values than the routine catalog across the region, due to improved sensitivity to small earthquakes. Although the improvement is region-wide, the degree of enhanced detectability varies spatially with station density and phase picking performance. Our analysis demonstrates clear advantages of the machine learning based catalog, but the level of improvement is not uniform and will vary with time. We observe greater variability in station-level detection capability for the machine learning based approach compared to the routine catalog. This implies that machine learning based monitoring may require region- or station-specific retraining or fine-tuning to achieve consistent performance \citep{liu2025evaluating, tan2025clearer, zhong2024deep}.

The analysis also identifies limitations in both the catalog and the framework. The method assumes that all detections from both catalogs are correct, but some of them, especially in the ML catalog \citep{suarez2025pervasive, park2023mitigation, park2025reducing}, may be false-positive picks that coincidentally align with an associated event and would act to bias the logistic model. This effect would artificially increase the estimated phase detection probabilities. In addition, fitting of the station-level probability model contains more uncertainty when pick quality varies or when events occupy a narrow magnitude-distance range. Future work could incorporate pick-level uncertainty and station-level detection probability uncertainty to evaluate how spatial detectability and its uncertainty evolve for machine learning based earthquake monitoring.

\section{Conclusions}
We developed a logistic based probability based magnitude of completeness (PMC) framework that quantifies station level detection probabilities and propagates them into spatial magnitude of completeness fields, enabling direct comparison of detection capability between seismic catalogs. Applied to the 2016 to 2017 Central Italy sequence, the results show that the machine learning based catalog consistently lowers completeness magnitudes relative to the routine catalog, while exhibiting larger variability in station level detection probability. This combined station level and spatial perspective identifies where machine learning enhances detectability and where further fine tuning is required, providing a quantitative basis for evaluating catalog performance and guiding seismic network design.

\section*{Acknowledgments}
We thank Yen Joe Tan for providing the ML catalog. This work was supported by the Department of Energy (Basic Energy Sciences; Award Number DE-SC0026319). 



\section*{Declarations}
\textbf{Conflict of Interest}  
The authors declare no conflict of interest.

\bibliographystyle{plainnat} 
\bibliography{references}  

\end{document}